\documentclass[12pt,draftnofoot,onecolumn]{IEEEtran_icc}%
\usepackage{amsmath,amssymb,amsthm}
\usepackage{graphicx,color,enumerate,array}
\usepackage{algpseudocode}
\usepackage{algorithm}
\usepackage{setspace}
\DeclareMathOperator*{\argmax}{arg\,max}
\newtheorem{theorem}{Theorem}

\newtheorem{lemma}[theorem]{Lemma}
\newtheorem{mydef}{Definition}

\newcommand{\mc}[1]{\mathcal{#1}}
\newcommand{\fis}[3]{\frac{1+\sum_{#2\neq #1}|h_{#1#2}|^2 #3_{#2}(h)}{|h_{#1#1}|^2}}
\newcommand{\fij}[3]{\frac{1+\sum_{#2}|h_{#1#2}|^2 #3_{#2}(h)}{|h_{#1#1}|^2}}

\title{Distributed Algorithms for Complete and Partial Information Games on Interference Channels}
\author{\IEEEauthorblockN{Krishna Chaitanya A, Utpal Mukherji, and Vinod Sharma\footnote{Part of this paper was presented in RAWNET Workshop, International Symposium on Modeling and Optimization in Mobile, Ad Hoc and Wireless Networks (WiOpt) 2015, Mumbai, India, May 2015.}}\\
\IEEEauthorblockA{Department of ECE, Indian Institute of Science, Bangalore-560012 \\ Email: $\lbrace$akc, utpal, vinod$\rbrace$ @ece.iisc.ernet.in}
}
\date{}
\begin{document}
\maketitle
\begin{abstract}
We consider a Gaussian interference channel with independent direct and cross link channel gains, each of which is independent and identically distributed across time. Each transmitter-receiver user pair aims to maximize its long-term average transmission rate subject to an average power constraint.  We formulate a stochastic game for this system in three different scenarios.  First, we assume that each user knows all direct and cross link channel gains.  Later, we assume that each user knows channel gains of only the links that are incident on its receiver.  Lastly, we assume that each user knows only its own direct link channel gain.  In all cases, we formulate the problem of finding a Nash equilibrium (NE) as a variational inequality (VI) problem.  We present a novel heuristic for solving a VI.  We use this heuristic to solve for a NE of power allocation games with partial information.  We also present a lower bound on the utility for each user at any NE in the case of the games with partial information.  We obtain this lower bound using a water-filling like power allocation that requires only knowledge of the distribution of a user's own channel gains and average power constraints of all the users.  We also provide a distributed algorithm to compute Pareto optimal solutions for the proposed games.  Finally, we use Bayesian learning to obtain an algorithm that converges to an $\epsilon$-Nash equilibrium for the incomplete information game with direct link channel gain knowledge only without requiring the knowledge of the power policies of the other users.
\end{abstract}
\section{Introduction}
Power allocation problem on interference channels is modeled in game theoretic framework and has been widely studied \cite{palomar}-\cite{rate_con}.  Most of the existing literature considered parallel Gaussian interference channels.  Nash equilibrium (NE) and Pareto optimal points are the main solutions obtained for the power allocation games.  While each user aiming to maximize its rate of transmission, for single antenna systems, NE is obtained in \cite{palomar} under certain conditions on the channel gains that also guarantee uniqueness.  Under these conditions the water-filling mapping is contraction map.  These results are extended to multi-antenna systems in \cite{MIMO_IWF}.  In the presence of multiple NE, an algorithm is proposed in \cite{VI} to find a NE that minimizes the total interference at all users among the NE.\par
An online algorithm to reach a NE for parallel Gaussian channels is presented in \cite{stochastic} when the channel gains are fixed but not known to the users.  Its convergence is also proved.\par
The power allocation problem on parallel Gaussian interference channels that minimize the total power subject to rate constraints for each user is considered in \cite{pareto}, \cite{d_PA}, and \cite{rate_con}.  NE is obtained under certain sufficient conditions in \cite{d_PA}.  Sequential and simultaneous iterative water-filling algorithms are proposed in \cite{rate_con} to find a NE.  Sufficient conditions for convergence of these algorithms are also studied.  Pareto optimal solutions are obtained by a decentralized iterative algorithm in \cite{pareto} assuming finite number of power levels for each user.\par
In \cite{PA} we consider a Gaussian interference channel with fast fading channel gains whose distributions are known to all the users.  We consider power allocation in a non-game-theoretic framework, and provide other references for such a set up.  In \cite{PA}, we have proposed a centralized algorithm for finding the Pareto points that maximize the average sum rate, when the receivers have knowledge of all the channel gains and decode the messages from strong and very strong interferers instead of treating them as noise as done in all the above references.\par
In this paper, we consider a stochastic game over additive Gaussian interference channels, where the users want to maximize their long term average rate and have long term average power constraints (for potential advantages of this over one shot optimization considered in the above references, see \cite{goldsmith}, \cite{vsharma}).  For this system we obtain existence of a NE and also develop a heuristic algorithm to find a NE under more general channel conditions for the complete information game.\par
We also consider the much more realistic situation when a user knows only its own channel gains, whereas the above mentioned literature considers the problem when each user knows all the channel gains in the system.  We consider two different partial information games.  In the first partial information game, each transmitter is assumed to have knowledge of the channel gains of the links that are incident on its corresponding receiver from all the transmitters.  Later, in the other game, we assume that each transmitter has knowledge of its direct link channel gain only.  For both the partial information games, we find a NE using the heuristic algorithm developed in the paper.\par
In each partial information game, we also present a lower bound on the average rate of each user at any Nash equilibrium.  This lower bound can be obtained by a user using a water-filling like, easy to compute power allocation, that can be evaluated with the knowledge of the distribution of its own channel gains and of the average power constraints of all the users. \par
We present a distributed algorithm to compute Pareto optimal and Nash bargaining solutions for all the three proposed games.  We obtain Pareto optimal points by maximizing the weighted sum of the uitlities (rates of transmission) of the all users.\par
Throughout, each user requires the knowledge of the channel statics and the power policies of other users.  Later we relax this assumption and use Bayesian learning to compute $\epsilon$-Nash equilibrium of the game in which only direct link channel gain is known at the corresponding transmitter.  But in this case, we consider finite strategy set, i.e., finite power levels rather than a continuum of powers considered before.\par
The paper is organized as follows. In Section \ref{sys_model}, we present the system model and the three stochastic game formulations.  Section \ref{one} reformulates the complete information stochastic game as an affine variational inequality problem.  In Section \ref{gen_vi}, we propose the heuristic algorithm to solve the formulated variational inequality under more general conditions.  In Section \ref{incomplete} we use this algorithm to obtain a NE when users have only partial information about the channel gains.  Pareto optimal and Nash bargaining solutions are discussed in Sections \ref{pareto}, \ref{nb} respectively and finally we apply Bayesian learning in Section \ref{bl}.  We present numerical examples in Section \ref{ne}.  Section \ref{concl} concludes the paper.
\section{System model and Stochastic Game Formulations} \label{sys_model}
We consider a Gaussian wireless channel being shared by $N$ transmitter-receiver pairs. The time axis is slotted and all users' slots are synchronized.  The channel gains of each transmit-receive pair are constant during a slot and change independently from slot to slot.  These assumptions are usually made for this system \cite{palomar}, \cite{vsharma}.\par
Let $H_{ij}(k)$ be the random variable that represents channel gain from transmitter $j$ to receiver $i$ (for transmitter $i$, receiver $i$ is the intended receiver) in slot $k$.  The direct channel power gains $\vert H_{ii}(k)\vert^2 \in \mathcal{H}_d = \lbrace g_1^{(d)},g_2^{(d)},\dots,g_{n_1}^{(d)} \rbrace$ and the cross channel power gains $\vert H_{ij}(k)\vert^2 \in \mathcal{H}_c = \lbrace g_1^{(c)},g_2^{(c)},\dots,g_{n_2}^{(c)}\rbrace$ where $n_1$, and $n_2$ are arbitrary positive integers.  We assume that, $\{H_{ij}(k), k \geq 0 \}$ is an $i.i.d$ sequence with distribution $\pi_{ij}$ where $\pi_{ij} = \pi_d$ if $i=j$ and $\pi_{ij} = \pi_c$ if $i\neq j$ and $\pi_d$ and $\pi_c$ are probability distributions on $\mathcal{H}_d$ and $\mathcal{H}_c$ respectively.  We also assume that these sequences are independent of each other.\par
We denote $(H_{ij}(k), i,j = 1,\dots,N)$ by ${\bf H}(k)$ and its realization vector by $h(k)$ which takes values in $\mathcal{H}$, the set of all possible channel states.  The distribution of ${\bf H}(k)$ is denoted by $\pi$.  We call the channel gains $(H_{ij}(k), j = 1,\dots,N)$ from all the transmitters to the receiver $i$ an incident gain of user $i$ and denote by ${\bf H}_i(k)$ and its realization vector by $h_i(k)$ which takes values in $\mathcal{I}$, the set of all possible incident channel gains.  The distribution of ${\bf H}_i(k)$ is denoted by $\pi_I$.\par
Each user aims to operate at a power allocation that maximizes its long term average rate under an average power constraint.  Since their transmissions interfere with each other, affecting their transmission rates,  we model this scenario as a stochastic game.\par
We first assume complete channel knowledge at all transmitters and receivers.  If user $i$ uses power $P_i({\bf H}(k))$ in slot $k$, it gets rate $\text{log} \left(1+\Gamma_i \left(P \left({\bf H}(k) \right) \right) \right)$, where
\begin{equation}
\Gamma_i(P({\bf H}(k))) = \frac{\alpha_i |H_{ii}(k)|^2 P_i({\bf H}(k))}{1 + \sum_{j \neq i}|H_{ij}(k)|^2P_j({\bf H}(k))},
\end{equation}
$P({\bf H}(k)) = (P_1({\bf H}(k)),\dots,P_N({\bf H}(k)))$ and $\alpha_i$ is a constant that depends on the modulation and coding used by transmitter $i$ and we assume $\alpha_i = 1$ for all $i$.  The aim of each user $i$  is to choose a power policy to maximize its long term average rate
\begin{equation}
r_i({\bf P}_i,{\bf P}_{-i}) \triangleq \limsup\limits_{n \rightarrow \infty} \frac{1}{n} \sum_{k=1}^n \mathbb{E}[\text{log} \left(1+\Gamma_i \left(P \left({\bf H}(k) \right) \right) \right)],
\end{equation}
subject to average power constraint
\begin{equation}
\limsup\limits_{n \rightarrow \infty} \frac{1}{n} \sum_{k=1}^n \mathbb{E}[P_i({\bf H}(k))] \leq \overline{P}_i, \text{ for each } i,\label{avg_c}
\end{equation}
where ${\bf P}_{-i}$ denotes the power policies of all users except $i$.  We denote this game by $\mathcal{G}_A$.\par
We next assume that the $i$th transmitter-receiver pair has knowledge of its incident gains ${\bf H}_i$ only.  Then the rate of user $i$ is
\begin{equation}
r_i({\bf P}_i,{\bf P}_{-i}) \triangleq \limsup\limits_{n \rightarrow \infty} \frac{1}{n} \sum_{k=1}^n \mathbb{E}_{{\bf H}_i(k)} \left[\mathbb{E}_{{\bf H}_{-i}(k)}[\text{log} \left(1+ \Gamma_i(P({\bf H}_i(k),{\bf H}_{-i}(k)))\right)]\right],
\end{equation}
where $P_i({\bf H}(k))$ depends only on ${\bf H}_i(k)$ and $\mathbb{E}_X$ denotes expectation with respect to the distribution of $X$.  Each user maximizes its rate subject to (\ref{avg_c}).  We denote this game by $\mathcal{G}_I$.\par
We also consider a game assuming that each transmitter-receiver pair knows only its direct link gain $H_{ii}$.  This is the most realistic assumption since each receiver $i$ can estimate $H_{ii}$ and feed it back to transmitter $i$.  In this case, the rate of user $i$ is given by
\begin{equation}
r_i({\bf P}_i,{\bf P}_{-i}) \triangleq \limsup\limits_{n \rightarrow \infty} \frac{1}{n} \sum_{k=1}^n \mathbb{E}_{{\bf H}_{ii}(k)} \left[ \mathbb{E}_{{\bf H}_{-ii}(k)}[\text{log} \left(1+ \Gamma_i(P(H_{ii}(k),H_{-ii}(k)))\right)]\right], \label{r_d}
\end{equation}
where $P_i({\bf H}(k))$ is a function of $H_{ii}(k)$ only.  Here, $H_{-ii}$ denotes the channel gains of all other links in the interference channel except $H_{ii}$.  In this game, each user maximizes its rate (\ref{r_d}) under the average power constraint (\ref{avg_c}).  We denote this game by $\mathcal{G}_D$.\par
We address these problems as stochastic games with the set of feasible power policies of user $i$ denoted by $\mathcal{A}_i$ and its utility by $r_i$.  Let $\mathcal{A} = \Pi_{i=1}^{N} \mathcal{A}_i$.\par
We limit ourselves to stationary policies, i.e., the power policy for every user in slot $k$ depends only on the channel state $H(k)$ and not on $k$.  In fact now we can rewrite the optimization problem in $\mathcal{G}_A$ to find policy $P({\bf H})$ such that $r_i = \mathbb{E}_{\bf H}[\text{log} \left(1+\Gamma_i \left(P \left({\bf H}\right) \right) \right)]$ is maximized subject to $\mathbb{E}_{\bf H}\left[P_i({\bf H})\right] \leq \overline{P}_i$ for all $i$.  We express power policy of user $i$ by ${\bf P}_i = (P_i(h), h\in \mathcal{H})$, where transmitter $i$ transmits in channel state $h$ with power $P_i(h)$.  We denote the power profile of all users by ${\bf P} = ({\bf P}_1,\dots,{\bf P}_N)$. \par
In the rest of the paper, we prove existence of a Nash equilibrium for each of these games and provide algorithms to compute it.\par
\section{Variational Inequality Formulation} \label{one}
We denote our game by $\mathcal{G}_A = \big((\mathcal{A}_i)_{i=1}^N, (r_i)_{i=1}^N\big)$, where $r_i({\bf P}_i,{\bf P}_{-i}) = \mathbb{E}_{\bf H}[\text{log} \left(1+\Gamma_i \left(P \left({\bf H}\right) \right) \right)]$ and $\mathcal{A}_i = \{{\bf P}_i \in \mathbb{R}^N: \mathbb{E}_{\bf H}\left[P_i({\bf H})\right] \leq \overline{P}_i, P_i(h) \geq 0 \text{ for all } h \in \mathcal{H}\}$.
\begin{mydef}
A point ${\bf P}^*$ is a Nash Equilibrium (NE) of game $\mathcal{G}_A = \big((\mathcal{A}_i)_{i=1}^N, (r_i)_{i=1}^N\big)$ if for each user $i$
\begin{equation*}
r_i({\bf P}_i^*,{\bf P}_{-i}^*) \geq r_i({\bf P}_i,{\bf P}_{-i}^*) \text{ for all } {\bf P}_i \in \mathcal{A}_i.
\end{equation*}
\end{mydef}
We now state Debreu-Glicksberg-Fan theorem (\cite{BASAR}, page no. 69) on the existence of a pure strategy NE.
\begin{theorem}\label{dgf}
Given a non-cooperative game, if every strategy set $\mathcal{A}_i$ is compact and convex, $r_i(a_i,a_{-i})$ is a continuous function in the profile of strategies ${\bf a} = (a_i,a_{-i}) \in \mathcal{A}$ and quasi-concave in $a_i$, then the game has atleast one pure-strategy Nash equilibrium. \qed
\end{theorem}
Existence of a pure NE for the strategic games $\mathcal{G}_A, \mathcal{G}_I$ and $\mathcal{G}_D$ follows from Theorem \ref{dgf}, since in our game $r_i({\bf P}_i,{\bf P}_{-i})$ is a continuous function in the profile of strategies ${\bf P} = ({\bf P}_i,{\bf P}_{-i}) \in \mathcal{A}$ and concave in ${\bf P}_i$ for $\mathcal{G}_A, \mathcal{G}_I$ and $\mathcal{G}_D$.  Also, $\mathcal{A}_i$ is compact and convex for each $i$.
\begin{mydef}
  The best-response of user $i$ is a function $BR_i : \mathcal{A}_{-i} \rightarrow \mathcal{A}_i$ such that $BR_i({\bf P}_{-i})$ maximizes $r_i({\bf P}_i, {\bf P}_{-i})$, subject to ${\bf P}_i \in \mathcal{A}_i$.   
\end{mydef}
A Nash equilibrium is a fixed point of the best-response function.  In the following we provide algorithms to obtain this fixed point for $\mathcal{G}_A$.  In Section \ref{incomplete} we will consider $\mathcal{G}_I$ and $\mathcal{G}_D$.  Given other users' power profile ${\bf P}_{-i}$, we use Lagrange method to evaluate the best response of user $i$.  The Lagrangian function is defined by
\begin{equation*}
\mathcal{L}_i({\bf P}_i,{\bf P}_{-i}) = r_i({\bf P}_i, {\bf P}_{-i}) + \mu_i( \overline{P}_i - \mathbb{E}_{\bf H}\left[P_i({\bf H})\right]).
\end{equation*}
To maximize $\mathcal{L}_i({\bf P}_i,{\bf P}_{-i})$, we solve for ${\bf P}_i$ such that $\frac{\partial \mathcal{L}_i}{\partial {\bf P}_i(h)} = 0$ for each $h \in \mathcal{H}$.  Thus, the component of the best response of user $i$, ${\bf BR}_i({\bf P}_{-i})$ corresponding to channel state $h$ is given by
\begin{equation}
BR_i({\bf P}_{-i};h) = \text{max}\left\{0, \lambda_i({\bf P}_{-i}) - \frac{(1+\sum_{j\neq i}\vert h_{ij}\vert^2 P_j(h))}{\vert h_{ii}\vert^2}\right\}, \label{wf}
\end{equation}
where $\lambda_i({\bf P}_{-i}) = \frac{1}{\mu_i({\bf P}_{-i})}$ is chosen such that the average power constraint is satisfied.\par
\hspace{0.25cm} It is easy to observe that the best-response of user $i$ to a given strategy of other users is water-filling on ${\bf f}_i({\bf P}_{-i}) = (f_i({\bf P}_{-i};h),h \in \mathcal{H})$ where 
\begin{equation}
f_i({\bf P}_{-i};h) = -\frac{(1+\sum_{j\neq i}\vert h_{ij}\vert^2 P_j(h))}{\vert h_{ii}\vert^2}.
\end{equation}
For this reason, we represent the best-response of user $i$ by ${\bf WF}_i({\bf P}_{-i})$.  The notation used for the overall best-response ${\bf WF}({\bf P}) = ({\bf WF}(P(h)), h \in \mathcal{H})$, where ${\bf WF}(P(h)) = (WF_1({\bf P}_{-1};h),\dots,WF_N({\bf P}_{-N};h))$ and $WF_i({\bf P}_{-i};h)$ is as defined in (\ref{wf}).  We use ${\bf WF}_i({\bf P}_{-i}) = (WF_i({\bf P}_{-i};h), h \in \mathcal{H})$.\par 
It is observed in \cite{palomar} that the best-response ${\bf WF}_i({\bf P}_{-i})$ is also the solution of the optimization problem
\begin{eqnarray}
&\text{ minimize } & \left\Vert {\bf P}_i  - {\bf f}_i({\bf P}_{-i})\right\Vert^2,  \nonumber \\
&\text{ subject to } & {\bf P}_i \in \mathcal{A}_i. \label{opt_proj}
\end{eqnarray}
As a result we can interpret the best-response as the projection of $(f_{i,1}({\bf P}_{-i}),\dots,f_{i,N}({\bf P}_{-i}))$ on to $\mathcal{A}_i$.  We denote the projection of $x$ on to $\mathcal{A}_i$ by $\Pi_{\mathcal{A}_i}(x)$.  We consider (\ref{opt_proj}), as a game in which every user minimizes its cost function $\left\Vert {\bf P}_i  - {\bf f}_i({\bf P}_{-i})\right\Vert^2$ with strategy set of user $i$ being $\mathcal{A}_i$.  We denote this game by $\mathcal{G}_A^{\prime}$.  This game has the same set of NEs as $\mathcal{G}_A$ because the best responses of these two games are equal. \par
The theory of variational inequalities offers various algorithms to find NE of a given game \cite{Pang}.  A variational inequality problem denoted by $VI(K,F)$ is defined as follows.
\begin{mydef}
Let $K \subset \mathbb{R}^n$ be a closed and convex set, and $F:K \to K$. The variational inequality problem $VI(K,F)$ is defined as the problem of finding $x \in K$ such that $$F(x)^T(y-x) \geq 0 \text{ for all } y \in K.$$
\end{mydef}
\begin{mydef}
We say that $VI(K,F)$ is  
\begin{itemize}
\item Monotone if $(F(x) -F(y))^T(x-y) \geq 0 \text{ for all } x,y \in K.$
\item Strictly monotone if $(F(x) -F(y))^T(x-y) > 0 \text{ for all } x,y \in K, x \neq y.$
\item Strongly monotone if there exists an $\epsilon >0 $ such that $(F(x) -F(y))^T(x-y) \geq \epsilon \Vert x-y \Vert^2 \text{ for all } x,y \in K$.
\end{itemize}
\end{mydef}
We reformulate the Nash equilibrium problem at hand to an affine variational inequality problem.  We now formulate the variational inequality problem corresponding to the game $\mathcal{G}_A^{\prime}$.\par
We note that (\ref{opt_proj}) is a convex optimization problem. The necessary and sufficient condition for $x^*$ to be solution of the convex optimization problem (\cite{Comp}, page 210)
\begin{equation*}
\text{ minimize }  g(x), \text{ subject to } x \in X,
\end{equation*}
 where $g(x)$ is a convex function and $X$ is a convex set,  is
 \begin{equation*}
\nabla g(x^*)(y-x^*) \geq 0 \text{ for all } y \in X.
\end{equation*}\par
Thus, given ${\bf P}_{-i}$, we need ${\bf P}_i^*$ for user $i$ such that
\begin{equation}
\sum_{h \in \mathcal{H}} \left( P_i^*(h) + f_i({\bf P}_{-i};h) \right)\left( x_i(h) - P_{i}^*(h) \right)\geq 0,\label{ineq}
\end{equation}
for all ${\bf x}_{i} \in \mathcal{A}_i$.  We can rewrite it more compactly as, 
\begin{equation}
\left({\bf P}^* + \hat{h} + \hat{H}{\bf P}^* \right)^T \left(x - {\bf P}^*\right) \geq 0 \text{ for all } x \in \mathcal{A}, \label{cond_1}
\end{equation}
where $\hat{h}$ is a $N_1$-length block vector with $N_1 = \vert \mathcal{H}\vert$, the cardinality of $\mathcal{H}$, each block $\hat{h}(h), h \in \mathcal{H}$, is of length $N$ and is defined by $\hat{h}(h) = \left( \frac{1}{\vert h_{11} \vert^2}, \dots, \frac{1}{\vert h_{NN} \vert^2}\right)$ and  $\hat{H}$ is the block diagonal matrix $\hat{H} = \text{diag}\left\lbrace \hat{H}(h), h \in \mathcal{H} \right\rbrace$ with each block $\hat{H}(h)$ defined by
\begin{equation*}
  [\hat{H}(h)]_{ij} = \begin{cases}
    0 & \text{ if } i=j, \\
    \frac{\vert h_{ij} \vert^2}{\vert h_{ii} \vert^2}, & \text{ else. }
    \end{cases}
\end{equation*}
The characterization of Nash equilibrium in (\ref{cond_1}) corresponds to solving for ${\bf P}$ in the affine variational inequality problem $VI(\mathcal{A},F)$,
\begin{equation}
F({\bf P})^T \left(x - {\bf P}\right) \geq 0 \text{ for all } x \in \mathcal{A},\label{vi_complete}
\end{equation}
where $F({\bf P}) =  (I+ \hat{H}){\bf P} + \hat{h}$.\par
In \cite{NCC15}, we presented an algorithm to compute NE when $\tilde{H} = I+ \hat{H}$ is positive semidefinite.  In, \cite{NCC15}, we proved that $\tilde{H}$ being positive semidefinite is a weaker sufficient condition than the existing condition in \cite{palomar}.
\section{Algorithm to Solve Variational Inequality Under General Channel Conditions}\label{gen_vi}
In this section we aim to find a NE even if $\tilde{H}$ is not positive semidefinite.  For this, we present a heuristic to solve the $VI(\mathcal{A},F)$ in general. \par
We base our heuristic algorithm on the fact that a power allocation ${\bf P}^*$ is a solution of $VI(\mathcal{A},F)$ if and only if
\begin{equation}
{\bf P}^* = \Pi_{\mathcal{A}}\left( {\bf P}^* - \tau F({\bf P}^*)\right), \label{fp_eq}
\end{equation}
for any $\tau > 0$.  We prove this fact using a property of projection on a convex set that can be stated as follows (\cite{Pang}):
\begin{lemma}\label{lemma_proj}
Let $X \subset \mathbb{R}^n$ be a convex set.  The projection $\Pi(y)$ of $y \in \mathbb{R}^n$, is the unique element in $X$ such that
\begin{equation}
(\Pi(y)-y)^T(x-\Pi(y)) \geq 0, \text{ for all } x \in X.\label{prop_proj}
\end{equation} \qed
\end{lemma}
Let ${\bf P}^*$ satisfy (\ref{fp_eq}) for some $\tau > 0$. By the property of projection (\ref{prop_proj}), we have
\begin{equation}
\left(\Pi_{\mathcal{A}}\left( {\bf P}^* - \tau F({\bf P}^*)\right) - \left( {\bf P}^* - \tau F({\bf P}^*)\right)\right)^T \left({\bf Q}-\Pi_{\mathcal{A}}\left( {\bf P}^* - \tau F({\bf P}^*)\right)\right) \geq 0 \label{prop_fp}
\end{equation}
for all ${\bf Q} \in \mathcal{A}$.  Using (\ref{fp_eq}) in (\ref{prop_fp}), we have
\begin{equation*}
\left( {\bf P}^* - \left( {\bf P}^* - \tau F({\bf P}^*) \right) \right)^T \left( {\bf Q} - {\bf P}^*\right) \geq 0,
\end{equation*}
\begin{equation*}
\text{i.e., } \left( \tau F({\bf P}^*) \right)^T \left( {\bf Q} - {\bf P}^*\right) \geq 0.
\end{equation*}
Since $\tau > 0$, we have
\begin{equation}
 F({\bf P}^*)^T \left( {\bf Q} - {\bf P}^*\right) \geq 0, \text{ for all } {\bf Q} \in \mathcal{A}.\label{vi_form}
\end{equation}
Thus ${\bf P}^*$ solves the $VI(\mathcal{A},F)$.  Conversely, let ${\bf P}^*$ be a solution of the $VI(\mathcal{A},F)$.  Then we have relation (\ref{vi_form}), which can be rewritten as
\begin{equation*}
 \left({\bf P}^* - ({\bf P}^* - \tau F({\bf P}^*)\right)^T \left( {\bf Q} - {\bf P}^*\right) \geq 0, \text{ for all } {\bf Q} \in \mathcal{A},
\end{equation*}
for any $\tau > 0$.  Comparing with (\ref{prop_proj}), from Lemma \ref{lemma_proj} we have that (\ref{fp_eq}) holds.  Thus, ${\bf P}^*$ is a fixed point of the mapping $T({\bf P}) = \Pi_{\mathcal{A}}\left( {\bf P} - \tau F({\bf P})\right)$ for any $\tau > 0$.\par
We can interpret the mapping $T({\bf P})$ as a better response mapping for the optimization (\ref{opt_proj}).  Consider a fixed point ${\bf P}^*$ of the better response $T({\bf P})$.  Then ${\bf P}^*$ is a solution of the variational inequality (\ref{vi_complete}).  This implies that, given ${\bf P}_{-i}^*$, ${\bf P}_i^*$ is a local optimum of (\ref{opt_proj}) for all $i$.  Since the optimization (\ref{opt_proj}) is convex, ${\bf P}_i^*$ is also a global optimum.  Thus given ${\bf P}_{-i}^*$, ${\bf P}_i^*$ is best response for all $i$, and hence a fixed point of the better response function $T({\bf P})$ is also a NE.\par 
To find a fixed point of $T({\bf P})$, we reformulate the variational inequality problem as a non-convex optimization problem
\begin{eqnarray}
&\text{ minimize } & \Vert {\bf P} - \Pi_{\mathcal{A}}\left( {\bf P} - \tau F({\bf P})\right)\Vert^2, \nonumber \\
&\text{ subject to } & {\bf P} \in \mathcal{A}.\label{s_obj}
\end{eqnarray}
The feasible region $\mathcal{A}$ of ${\bf P}$, can be written as a Cartesian product of $\mathcal{A}_i$, for each $i$, as the constraints of each user are decoupled in power variables.  As a result, we can split the projection $\Pi_{\mathcal{A}}(.)$ into multiple projections $\Pi_{\mathcal{A}_i}(.)$ for each $i$, i.e., $\Pi_{\mathcal{A}}({\bf x}) = (\Pi_{\mathcal{A}_1}({\bf x}_1),\dots,\Pi_{\mathcal{A}_N}({\bf x}_N))$.  For each user $i$, the projection operation $\Pi_{\mathcal{A}_i}({\bf x}_i)$ takes the form 
\begin{equation}
\Pi_{\mathcal{A}_i}({\bf x}_i) = \left(\text{max}\left(0, x_i\left(h\right)-\lambda_i\right), h \in \mathcal{H}\right),\label{pro_form}
\end{equation}
 where $\lambda_i$ is chosen such that the average power constraint is satisfied.  Using (\ref{pro_form}), we rewrite the objective function in (\ref{s_obj}) with $\tau =1$ as
\begin{equation}
\Vert {\bf P} - \Pi_{\mathcal{A}}\left( {\bf P} - F({\bf P})\right)\Vert^2  = \sum_{h \in \mathcal{H},i} \left(P_i\left(h\right)- \text{max}\left\{0, -f_i({\bf P}_{-i};h)-\lambda_i\right\}\right)^2 \nonumber
\end{equation}
\begin{equation*}
= \sum_{h \in \mathcal{H},i} \left(\text{min}\left\{P_i(h), \fij{i}{j}{P}+\lambda_i\right\}\right)^2
\end{equation*}
\begin{equation}
= \sum_{h \in \mathcal{H},i} \left(\text{min}\left\{P_i(h), P_i(h) + f_i({\bf P}_{-i};h)+\lambda_i\right\}\right)^2 .\label{sim_obj}
\end{equation}
At a NE, the left side of equation (\ref{sim_obj}) is zero and hence each minimum term on the right side of the equation must be zero as well.  This happens, only if, for each $i$, 
\begin{equation*}
P_i(h) = \begin{cases}
  0, \text{ if } \fis{i}{j}{P}+\lambda_i > 0, \\
  -\fis{i}{j}{P}-\lambda_i, \text{ otherwise. }
\end{cases}
\end{equation*}
Here, the Lagrange multiplier $\lambda_i$ can be negative, as the projection satisfies the average power constraint with equality.  At a NE user $i$ will not transmit if the ratio of total interference plus noise to the direct link gain is more than some threshold.\par
We now propose a heuristic algorithm to find an optimizer of (\ref{s_obj}).  This algorithm consists of two phases.  Phase 1 attempts to find a better power allocation, using Picard iterations with the mapping $T({\bf P})$, that is close to a NE.  We use Phase 1 in Algorithm \ref{s_min} to get a good initial point for the steepest descent algorithm of Phase 2.  We will show in Section \ref{ne} that it indeed provides a good initial point for Phase 2.  In Phase 2, using the estimate obtained from Phase 1 as the initial point, the algorithm runs the steepest descent method to find a NE.  It is possible that the steepest descent algorithm may stop at a local minimum which is not a NE.  This is because of the non-convex nature of the optimization problem.  If the steepest descent method in Phase 2 terminates at a local minimum which is not a NE, we again invoke Phase 1 with this local minimum as the initial point and then go over to Phase 2.  We present the complete algorithm below as Algorithm \ref{s_min}.\par
\begin{algorithm}[h]
  \begin{algorithmic}
    \State Fix $\epsilon > 0, \delta > 0$ and a positive integer MAX 
    \State {\bf Phase 1} :  Initialization phase\\
    \State Initialize ${\bf P}_i^{(0)}$ for all $i = 1,\dots,N$.    
    \For {$n = 1 \to \text{MAX} $}
    \State $ {\bf P}^{(n)} = $ $T({\bf P}^{(n-1)})$\vspace{0.2cm}    
    \EndFor
    \State go to Phase 2.\\
    \State {\bf Phase 2} : Optimization phase\\
    \State Initialize $t = 1, {\bf P}^{(t)} = {\bf P}^{(MAX)}$,
    \Loop
    \State For each $i$, ${\bf P}_i^{(t+1)}$ = Steepest\_Descent($\tilde{{\bf P}}_i^{(t)},i$)
    \State where $\tilde{{\bf P}}_i^{(t)} = ({\bf P}_1^{(t+1)},\dots,{\bf P}_{i-1}^{(t+1)},{\bf P}_i^{(t)},\dots,{\bf P}_N^{(t)})$,
    \State ${\bf P}^{(t+1)} = ({\bf P}_1^{(t+1)},\dots,{\bf P}_N^{(t+1)})$,    
    \State Till $\Vert {\bf P}^{(t+1)} - T({\bf P}^{(t+1)}) \Vert < \epsilon$
    \If {$\Vert {\bf P}^{(t)} -{\bf P}^{(t+1)}\Vert < \delta $ and $\Vert {\bf P}^{(t+1)} - T({\bf P}^{(t+1)})\Vert > \epsilon$}
    \State Go to Phase 1 with ${\bf P}^{(0)} = {\bf P}^{(t+1)}$    
    \EndIf
    \State $t = t+1$.
    \EndLoop
    \Function{$\text{Steepest\_Descent}$}{${\bf P}^{(t)},i$}\\    
    \State $\triangledown f({\bf P}^{(t)}) = (\frac{\partial f({\bf P})}{\partial P_i(h)}\vert_{{\bf P} = {\bf P}^{(t)}}, h \in \mathcal{H})$ 
    \State where $f({\bf P}) = \Vert {\bf P} - T({\bf P}) \Vert^2$ 
    \For {$h \in \mathcal{H}$}    
    \State evaluate $\frac{\partial f({\bf P})}{\partial P_i(h)}\vert_{{\bf P} = {\bf P}^{(t)}}$ using derivative approximation
    \EndFor
    \State ${\bf P}_i^{(t+1)} = \Pi_{\mathcal{A}_i}({\bf P}_i^{(t)} - \gamma_t\triangledown f({\bf P}^{(t)}))$        
    \State return ${\bf P}_i^{(t+1)}$
    \EndFunction    
  \end{algorithmic}  
  \caption{Heuristic algorithm to find a Nash equilibrium}  
  \label{s_min}  
\end{algorithm}
\section{Partial Information games}\label{incomplete}
In the partial information games, unlike in complete information game, we can not write the problem of finding a NE as an affine variational inequality, because the best response is not water-filling and should be evaluated numerically.  But we can still formulate the problem of finding a NE as a non-affine VI.  In this section, we show that we can use Algorithm \ref{s_min} to find a NE even for these games.\par
\subsection{Game $\mathcal{G}_I$}
We first consider the game $\mathcal{G}_I$ and find its NE using Algorithm \ref{s_min}.  We write the variational inequality formulation of the NE problem.  For user $i$, the optimization at hand is 
\begin{eqnarray}
& \text{ maximize } & r_i^{(I)}, \nonumber \\
& \text{ subject to } & {\bf P}_i \in \mathcal{A}_i,\label{in_pr}
\end{eqnarray}
where $r_i^{(I)} = \sum_{h_i \in \mathcal{I}}\pi(h_i)\mathbb{E}\left[\text{log} \left(1+ \frac{ |h_{ii}|^2 P_i({\bf h}_i)}{1 + \sum_{j \neq i}|h_{ij}|^2P_j(H_j)}\right)\right]$.  The necessary and sufficient optimality conditions for the convex optimization problem (\ref{in_pr}) are
\begin{equation}
({\bf x}_i - {\bf P}_i^*)^T(-\triangledown_ir_i^{(I)}({\bf P}_i^*,{\bf P}_{-i})) \geq 0, \text{ for all }{\bf x}_i \in \mathcal{A}_i,\label{ns_in}
\end{equation}
where $\triangledown_ir_i^{(I)}({\bf P}_i^*,{\bf P}_{-i})$ is the gradient of $r_i^{(I)}$ with respect to power variables of user $i$.  Then ${\bf P}^*$ is a NE if and only if (\ref{ns_in}) is satisfied for all $i = 1,\dots,N$.
We can write the $N$ inequalities in (\ref{ns_in}) as
\begin{equation}
({\bf x} - {\bf P}^*)^TF({\bf P}^*) \geq 0, \text{ for all }{\bf x} \in \mathcal{A},\label{vi_in}
\end{equation}
where $F({\bf P}) = (-\triangledown_1r_1^{(I)}({\bf P}),\dots,-\triangledown_Nr_N^{(I)}({\bf P}))^T$.  Equation (\ref{vi_in}) is the required variational inequality characterization.  A solution of the variational inequality is a fixed point of the mapping $T_I({\bf P}) = \Pi_{\mathcal{A}}({\bf P} - \tau F({\bf P}))$, for any $\tau > 0$.  We use Algorithm \ref{s_min}, to find a fixed point of $T_I({\bf P})$ by replacing $T({\bf P})$ in Algorithm \ref{s_min} with $T_I({\bf P})$.\par
\subsection{Better response iteration}
\hspace{0.25cm} In this subsection, we interpret $T_I({\bf P})$ as a better response for each user.  For this, consider the optimization problem (\ref{in_pr}).  For this, using the gradient projection method, the update rule for power variables of user $i$ is 
\begin{equation}
{\bf P}_i^{(n+1)} = \Pi_{\mathcal{A}_i}({\bf P}_i^{(n)} + \tau \triangledown_ir_i^{(I)}({\bf P}^{(n)})).\label{btr}
\end{equation}
The gradient projection method ensures that for a given ${\bf P}_{-i}^{(n)}$, $$r_i^{(I)}({\bf P}_i^{(n+1)},{\bf P}_{-i}^{(n)}) \geq r_i^{(I)}({\bf P}_i^{(n)},{\bf P}_{-i}^{(n)}).$$  Therefore, we can interpret ${\bf P}_i^{(n+1)}$ as a better response to ${\bf P}_{-i}^{(n)}$ than ${\bf P}_i^{(n)}$.  As the feasible space $\mathcal{A} = \Pi_{i=1}^N \mathcal{A}_i$, we can combine the update rules of all users and write 
\begin{equation*}
{\bf P}^{(n+1)} = \Pi_{\mathcal{A}}({\bf P}^{(n)} - \tau F({\bf P}^{(n)})) = T_I({\bf P}^{(n)}).
\end{equation*}
Thus, the Phase $1$ of Algorithm \ref{s_min} is the iterated better response algorithm.\par
Consider a fixed point ${\bf P}^*$ of the better response $T_I({\bf P})$.  Then ${\bf P}^*$ is a solution of the variational inequality \ref{vi_in}.  This implies that, given ${\bf P}_{-i}^*$, ${\bf P}_i^*$ is a local optimum of (\ref{in_pr}) for all $i$.  Since the optimization (\ref{in_pr}) is convex, ${\bf P}_i^*$ is also a global optimum.  Thus given ${\bf P}_{-i}^*$, ${\bf P}_i^*$ is best response for all $i$, and hence a fixed point of the better response function is also a NE.  This gives further justification for Phase 1 of Algorithm \ref{s_min}.  Indeed we will show in the next section that in such a case Phase 1 often provides a NE for $\mathcal{G}_I$ and $\mathcal{G}_D$ (for which also Phase 1 provides a better response dynamics; see Section \ref{direct} below).
\subsection{Game $\mathcal{G}_D$}\label{direct}
We now consider the game $\mathcal{G}_D$ where each user $i$ has knowledge of only the corresponding direct link gain $H_{ii}$.  In this case also we can formulate the variational inequality characterization.  The variational inequality becomes
\begin{equation}
({\bf x} - {\bf P}^*)^TF_D({\bf P}^*) \geq 0, \text{ for all }{\bf x} \in \mathcal{A},\label{vi_d}
\end{equation}
where $F_D({\bf P}) = (-\triangledown_1r_1^{(D)}({\bf P}),\dots,-\triangledown_Nr_N^{(D)}({\bf P}))^T$, 
\begin{equation}  
r_i^{(D)} = \sum_{h_{ii}}\pi(h_{ii})\mathbb{E}\left[\text{log} \left(1+ \frac{ |h_{ii}|^2 P_i(h_{ii})}{1 + \sum_{j \neq i}|H_{ij}|^2P_j(H_j)}\right)\right].\label{rate_direct}
\end{equation}
We use Algorithm \ref{s_min} to solve the variational inequality (\ref{vi_d}) by finding fixed points of $T_D({\bf P}) = \Pi_{\mathcal{A}}({\bf P} - \tau F_D({\bf P}))$.  Also, one can show that as for $T_I$, $T_D$ provides a better response strategy.
\subsection{Lower bound}
In this subsection, we derive a lower bound on the average rate of each user at any NE.  This lower bound can be achieved at a water-filling like power allocation that can be computed with knowledge of only its own channel gain distribution and the average power constraint of all the users.\par
To compute a NE using Algorithm \ref{s_min}, each user needs to communicate its power variables to the other users in every iteration and should have knowledge of the distribution of the channel gains of all the users.  If any transmitter fails to receive power variables from other users, it can operate at the water-filling like power allocation that attains at least the lower bound derived in this section.  Other users can compute the NE of the game that is obtained by removing the user that failed to receive the power variables, but treating the interference from this user as a constant, fixed by its water-filling like power allocation.  We now derive the lower bound.
\subsubsection{For $\mathcal{G}_I$}
 In the computation of NE, each user $i$ is required to know the power profile ${\bf P}_{-i}$ of all other users.  We now give a lower bound on the utility $r_i^{(I)}$ of user $i$ that does not depend on other users' power profiles.\par
We can easily prove that the function inside the expectation in $r_i^{(I)}$ is a convex function of ${\bf P}_j(h_j)$ for fixed ${\bf P}_i(h_i)$ using the fact that (\cite{boyd}) a function $f:\mathcal{K} \subseteq \mathbb{R}^n \rightarrow \mathbb{R}$ is convex if and only if $$\frac{d^2 f({\bf x}+t {\bf y})}{dt^2} \geq 0,$$ for all ${\bf x},{\bf y} \in \mathcal{K}$ and $t \in \mathbb{R}$ is such that ${\bf x}+t {\bf y} \in \mathcal{K}$. Then by Jensen's inequality to the inner expectation in $r_i^{(I)}$,
\begin{eqnarray}
r_i^{(I)} & = &\sum_{h_i \in \mathcal{I}}\pi(h_i)\mathbb{E}\left[\text{log} \left(1+ \frac{ |h_{ii}|^2 P_i({\bf h}_i)}{1 + \sum_{j \neq i}|h_{ij}|^2P_j(H_j)}\right)\right] \nonumber \\
& \geq & \sum_{h_i \in \mathcal{I}}\pi(h_i) \text{log} \left(1+ \frac{ |h_{ii}|^2 P_i({\bf h}_i)}{1 + \sum_{j \neq i}|h_{ij}|^2\mathbb{E}[P_j(H_j)]}\right)\nonumber \\
& = & \sum_{h_i \in \mathcal{I}}\pi(h_i) \text{log} \left(1+ \frac{ |h_{ii}|^2 P_i({\bf h}_i)}{1 + \sum_{j \neq i}|h_{ij}|^2\overline{P_j}}\right). \label{in_lb}
\end{eqnarray}
The above lower bound $r_{i,LB}^{(I)}({\bf P}_i)$ of $r_i^{(I)}({\bf P}_i,{\bf P}_{-i})$ does not depend on the power profile of users other than $i$.  We can choose a power allocation ${\bf P}_i$ of user $i$ that maximizes $r_{i,LB}^{(I)}({\bf P}_i)$.  It is the water-filling solution given by
\begin{equation*}
P_i(h_i) = \text{ max }\left\{0, \lambda_i - \frac{1 + \sum_{j \neq i}|h_{ij}|^2\overline{P_j}}{|h_{ii}|^2} \right\}.
\end{equation*}
Let ${\bf P}^* = ({\bf P}_i^*, {\bf P}_{-i}^*)$ be a NE, and let ${\bf P}_i^{\dagger}$ be the maximizer for the lower bound $r_{i,LB}^{(I)}({\bf P}_i)$.  Then, $r_{i}^{(I)}({\bf P}_i^*,{\bf P}_{-i}^*) \geq r_{i}^{(I)}({\bf P}_i, {\bf P}_{-i}^*) \text{ for all } {\bf P}_i \in \mathcal{A}_i$, in particular for ${\bf P}_i = {\bf P}_i^{\dagger}$.  Thus, $r_{i}^{(I)}({\bf P}_i^*,{\bf P}_{-i}^*) \geq r_{i}^{(I)}({\bf P}_i^{\dagger}, {\bf P}_{-i}^*)$. But, $r_{i}^{(I)}({\bf P}_i^{\dagger}, {\bf P}_{-i}^*) \geq r_{i,LB}^{(I)}({\bf P}_i^{\dagger})$.  Therefore, $r_{i}^{(I)}({\bf P}_i^*,{\bf P}_{-i}^*) \geq r_{i,LB}^{(I)}({\bf P}_i^{\dagger})$.  But, in general it may not hold that $r_{i}^{(I)}({\bf P}_i^*,{\bf P}_{-i}^*) \geq r_{i}^{(I)}({\bf P}_i^{\dagger},{\bf P}_{-i}^{\dagger})$.\par
\subsubsection{For $\mathcal{G}_D$}
We can also derive a lower bound on $r_i^{(D)}$ using convexity and Jensen's inequality as in (\ref{in_lb}).  In the case of $\mathcal{G}_D$, we have
\begin{equation*}
r_i^{(D)} \geq \sum_{h_{ii}}\pi(h_{ii}) \text{log} \left(1+ \frac{ |h_{ii}|^2 P_i(h_{ii})}{1 + \sum_{j \neq i}\mathbb{E}[|H_{ij}|^2]\overline{P_j}}\right).
\end{equation*}
The optimal solution for maximizing the lower bound is the water-filling solution
\begin{equation*}
  P_i(h_{ii}) = \text{ max }\left\{0, \lambda_i - \frac{1 + \sum_{j \neq i}\mathbb{E}[|H_{ij}|^2]\overline{P_j}}{|h_{ii}|^2} \right\}.
\end{equation*}
\section{Pareto Optimal Solutions}\label{pareto}
In this section, we consider Pareto optimal solutions to the game $\mathcal{G}$.  A power allocation ${\bf P}^{*}$ is Pareto optimal if there does not exist a power allocation ${\bf P}$ such that $r_i({\bf P}_i,{\bf P}_{-i}) \geq r_i({\bf P}_i^{*},{\bf P}_{-i}^{*})$ for all $i = 1,\dots,N$ with at least one strict inequality.  It is well-known that the solution of optimization problem, 
\begin{equation}
\text{ max } \sum_{i=1}^N w_i r_i({\bf P}_i,{\bf P}_{-i}),\text{ such that } {\bf P}_i \in \mathcal{A}_i \text{ for all } i, \label{ws_pareto}
\end{equation}
with $w_i > 0$, is Pareto optimal.  Thus, since $\mathcal{A}$ is compact and $r_i$ are continuous, a Pareto point exists for our problem.  We apply the weighted-sum optimization (\ref{ws_pareto}) to the game $\mathcal{G}$ to find a Pareto-optimal power allocation. \par
To solve the non-convex optimization problem in a distributed way, we employ augmented Lagrangian method \cite{lno} and solve for the stationary points using the algorithm in \cite{distributed}.  We present the resulting algorithm to find the Pareto power allocation in Algorithm \ref{algo_pareto}.  Define the augmented Lagrangian as 
\begin{multline*}
\mathcal{L}({\bf P},{\bf \lambda}) =  \sum_{i=1}^N w_i r_i({\bf P}_i,{\bf P}_{-i}) + \sum_{i=1}^N \lambda_i(\overline{P}_i-\sum_{h \in \mathcal{H}}\pi(h) P_i(h)) + c \sum_{i}(\overline{P}_i-\sum_{h \in \mathcal{H}}\pi(h) P_i(h))^2.
\end{multline*}
\begin{algorithm}
  \begin{algorithmic}
    \State Initialize $\lambda_i^{(1)},{\bf P}_i^{(0)}$ for all $i = 1,\dots,N$.
    \For {$n = 1 \to \infty $}
    \State $ {\bf P}^{(n)} = $ $\text{Steepest\_Ascent}(\lambda^{(n)},{\bf P}^{(n-1)})$\vspace{0.2cm}
    \If {$|\overline{P}_i-\sum_{h}\pi(h) P_i^{(n)}(h)| < \epsilon \text{ for all } i = 1,\dots,N $}
    \State break
    \Else
    \State $\lambda_i^{(n+1)} = \lambda_i^{(n)} - \alpha (\overline{P}_i-\sum_{h}\pi(h) P_i^{(n)}(h))$    
    \State $n = n+1$
    \EndIf
    \EndFor    
    \Function{$\text{Steepest\_Ascent}$}{$\lambda,{\bf P}$}
    \State Fix $\delta, \epsilon$
    \State Initialize $t = 1, {\bf P}^{(t)} = {\bf P}$.
    \Loop
    \For {$i=1 \to N$}
    \State player $i$ updates his power variables as \\ ${\bf Q}_i = {\bf P}_i^{(t)} + \delta \triangledown_i \mathcal{L}({\bf P}_i^{(t)},{\bf P}_{-i}^{(t)},\lambda)$
    \EndFor
    \State Choose ${\bf P}^{(t+1)}$ as 
    \State $i^* = \argmax_i \mathcal{L}({\bf Q}_i,{\bf P}_{-i}^{(t)},\lambda) - \mathcal{L}({\bf P}_i^{(t)},{\bf P}_{-i}^{(t)},\lambda)$ 
    \State ${\bf P}^{(t+1)} = ({\bf Q}_{i^*},{\bf P}_{-i^*}^{(t)}) $
    \State $t = t+1$.
    \State Till $\Vert \triangledown_i\mathcal{L}({\bf P}_i^{(t)},{\bf P}_{-i}^{(t)},\lambda)\Vert_2  < \epsilon$ for each $i$.
    \EndLoop
    \State return ${\bf P}^{(t)}$
    \EndFunction
  \end{algorithmic}
  \caption{Augmented Lagrangian method to find Pareto optimal power allocation}
  \label{algo_pareto}
\end{algorithm}
We denote the gradient of $\mathcal{L}({\bf P},{\bf \lambda})$ with respect to power variables of player $i$ by $\triangledown_i \mathcal{L}({\bf P}_i,{\bf P}_{-i},\lambda)$.  In Algorithm \ref{algo_pareto}, the step sizes $\alpha,\delta$ are chosen sufficiently small.  Convergence of the steepest ascent function in Algorithm \ref{algo_pareto} is proved in \cite{distributed}. \par
In a similar way, we can find Pareto optimal points for partial information games $\mathcal{G}_I$ and $\mathcal{G}_D$ by solving the optimization (\ref{ws_pareto}) with $r_i$ replaced by $r_i^{(I)}$ and $r_i^{(D)}$ respectively.  We can extend the algorithm to compute Pareto optimal power allocation for games $\mathcal{G}_I$ and $\mathcal{G}_D$ by appropriately redefining the augmented Lagrangian as
\begin{multline*}
\mathcal{L}^{(I)}({\bf P},{\bf \lambda}) =  \sum_{i=1}^N w_i r_i^{(I)}({\bf P}_i,{\bf P}_{-i}) + \sum_{i=1}^N \lambda_i(\overline{P}_i-\sum_{h \in \mathcal{H}}\pi(h) P_i(h)) + c \sum_{i}(\overline{P}_i-\sum_{h \in \mathcal{H}}\pi(h) P_i(h))^2.
\end{multline*}
for game $\mathcal{G}_I$ and 
\begin{multline*}
\mathcal{L}^{(D)}({\bf P},{\bf \lambda}) =  \sum_{i=1}^N w_i r_i^{(D)}({\bf P}_i,{\bf P}_{-i}) + \sum_{i=1}^N \lambda_i(\overline{P}_i-\sum_{h \in \mathcal{H}}\pi(h) P_i(h)) + c \sum_{i}(\overline{P}_i-\sum_{h \in \mathcal{H}}\pi(h) P_i(h))^2.
\end{multline*}
for game $\mathcal{G}_D$.
Since this is a nonconvex optimization problem, Algorithm \ref{algo_pareto} converges to a local Pareto point (\cite{MOP}) depending on the initial power allocation.  We can get better local Pareto points by initializing the algorithm from different power allocations and choosing the Pareto point which gives the best sum rate among the ones obtained.  We consider this in our illustrative examples.
\section{Nash Bargaining}\label{nb}
In general, Pareto optimal points do not guarantee fairness among users, i.e., Algorithm \ref{algo_pareto} may converge to a Pareto point such that a particular user at the Pareto point receives higher rate while another user at the same Pareto point receives arbitrarily small rate.  In this section we consider Nash bargaining solutions which are also Pareto optimal solutions but guarantee fairness.\par
In Nash bargaining, we specify a disagreement outcome that specifies utility of each user that it receives by playing the disagreement strategy if the utility received in the bargaining outcome is less than that received in the disagreement outcome for any user. \par
Thus, by choosing the disagreement outcomes appropriately, the users can ensure certain fairness.  The Nash bargaining solutions are Pareto optimal and also satisfy certain natural axioms \cite{nash_b}.  It is shown in \cite{nash_b} that for a two player game, there exists a unique bargaining solution (if the feasible region is nonempty) that satisfies the axioms stated above and it is given by the solution of the optimization problem
\begin{eqnarray}
\text{maximize } & & (r_1-d_1)(r_2-d_2), \nonumber \\
\text{subject to } & & r_i \geq d_i, i = 1,2, \nonumber\\
& & (r_1,r_2) \in \mc{R}.
\end{eqnarray}
For an N-user Nash bargaining problem, this result can be extended and the solution of an N-user bargaining problem is the solution of the optimization problem
\begin{eqnarray}
\text{maximize } & & \Pi_{i=1}^N (r_i-d_i), \nonumber \\
\text{subject to } & & r_i \geq d_i, i = 1,\dots,N \nonumber\\
& & (r_1,\dots,r_N) \in \mc{R}.\label{nb_n}
\end{eqnarray}\par
A Nash bargaining solution is also related to proportional fairness, another fairness concept commonly used in communication literature.  A utility vector $r^* \in \mc{R}$ is said to be \emph{proportionally fair} if for any other feasible vector $r \in \mc{R}$, for each ${\bf P}$, the aggregate proportional change $r_k-r^*_k/r_k^*$ is non-positive \cite{prop_fair}.  If the set $\mc{R}$ is convex, then Nash bargaining and proportional fairness are equivalent \cite{prop_fair}.\par
A major issue in finding a solution of a bargaining problem is choosing the disagreement outcome.  It is more common to consider an equilibrium point as a disagreement outcome.  In our problem we can consider the utility vector at a NE as the disagreement outcome.  We can also choose $d_i = 0$ for each $i$.  For our numerical evaluations we have chosen the disagreement outcome to be a zero vector.  To find the bargaining solution, i.e., to solve the optimization problem (\ref{nb_n}), we use the algorithm of Section \ref{pareto} used to find a Pareto optimal point but with the objective function
\begin{equation*}
 \Pi_{i=1}^N (r_i({\bf P})-d_i).
\end{equation*}
In Section \ref{ne}, we present a Nash bargaining solution for the numerical examples we consider and observe that the Nash bargaining solution obtained is a Pareto optimal point which provides fairness among the users.
\section{Bayesian Learning}\label{bl}
In Section \ref{gen_vi}, we discussed a heuristic algorithm to find a NE when the matrix $\tilde{H}$ is not positive semidefinite.  Even though the heuristic can be used to compute a NE, we do not have a proof of its convergence to a NE.  In this section, we use Bayesian learning that guarantees convergence to an $\epsilon$-Nash equilibrium of the partial information game $\mathcal{G}_D$ where $\epsilon >0$ can be chosen arbitrarily small.  We first define an $\epsilon$-Nash equilibrium.
\begin{mydef}
A point ${\bf P}^*$ is an $\epsilon$-Nash Equilibrium ($\epsilon$-NE) of game $\mathcal{G}_D = \big((\mathcal{A}_i)_{i=1}^N, (r_i^{D})_{i=1}^N\big)$ if for each user $i$
\begin{equation*}
r_i^D({\bf P}_i^*,{\bf P}_{-i}^*) \geq r_i^D({\bf P}_i,{\bf P}_{-i}^*)-\epsilon \text{ for all } {\bf P}_i \in \mathcal{A}_i.
\end{equation*}
\end{mydef}\par
Bayesian learning to find a NE for finite games is introduced in \cite{b_learn}.  In finite games, there are a finite number of users and the strategy set $\mathcal{A}_i$ of user $i$, is finite for all $i$.  Let the probability distribution $\phi_i$ on $\mathcal{A}_i$ be the strategy of user $i$, i.e., for each $k_i \in \mathcal{A}_i$, $\phi_i(k_i)$ denotes the probability that user $i$ plays the action $k_i$ under the strategy $\phi_i$.  In the model considered in \cite{b_learn}, a static game is played repeatedly for an infinite horizon and users update their strategies $\phi_i$ each time the game is played.  No user has knowledge about the opponents' strategy $\phi_{-i} = (\phi_1,\dots,\phi_{i-1},\phi_{i+1},\dots,\phi_{N})$.  But each user is provided with the actions chosen by the opponents in a time slot at the end of that time slot.  Let $k_i^t$ be the action chosen by user $i$ in time slot $t$. \par
In Bayesian learning, each user $i$ has a belief $\phi_j^i$ about strategy of user $j$, for each $j$ with $\phi_i^i = \phi_i$.  Each user, after every time slot $t$, finds the posterior belief on the opponents' strategies from the prior beliefs using Bayesian update rule, after observing the opponents actions in time slot $t$.  After finding the posterior beliefs $\phi^i_{-i}$, each user chooses a strategy that maximizes its utility, assuming that all the opponents follow their beliefs.  Following this procedure, it is shown in \cite{b_learn} that the posterior beliefs of all players converge to a $\epsilon$-NE.\par
We adapt this procedure to find a $\epsilon$-NE of our game $\mathcal{G}_D$.  For this, we consider finite power levels at which a user can transmit.  Let $\mathcal{P}_i = \{p^{(1)}_i,\dots,p^{(m)}_i\}$ be the set of power levels at which user $i$ can transmit.  These power levels can be different for different users.  Then the strategy set $\mathcal{A}_i$ of user $i$ with average power constraint is
\begin{equation}
\mathcal{A}^{(i)} = \left\{{\bf P}^{(i)}=(P^{(1)}_{i},\dots,P^{(n_i)}_{i})|P^{(l)}_i \in \{p_i^{(1)},\dots,p_i^{(m)}\},\right. \left. \sum_{l=1}^{n_i}\pi_{i}(l)P^{(l)}_i \leq \overline{P}_i \right\},
\end{equation}
where $\pi_i(l)$ is the probability of occurrence of $l^{th}$ channel state.  The strategy set of each user is finite.  To use the traditional Bayesian learning, each user needs to know the action chosen by the other users.  In general, a transmitter can not observe powers used by the other transmitters.  The purpose of knowing the actions of other users in Bayesian learning is to learn the strategies of other players and each user finds its best response with respect to the learned strategies of other users.  In our problem, for each user to find its best response to a given strategy of the other users, it is enough to know the interference level that is experienced by its corresponding receiver.  As the receiver can feedback the interference it has seen in a slot by the end of that slot to its transmitter, it is enough to learn the distribution of the interference rather than the strategy of the other users.  Hence, each user has belief on the distribution of the interference rather than having a belief on the strategies of opponents.  Each user updates its belief on the distribution on the interference using the Bayesian update with the help of feedback from its receiver.\par
Let $\mathcal{I}_i$ denote the set of possible interference levels for user $i$.  We denote the belief of user $i$ about the distribution of interference experienced at its receiver by $\phi^i$.  With respect to this belief $\phi^i$, user $i$ finds the best response $\phi_i$ and chooses an action according to the best response.  Please note that $\phi^i$ is a probability mass function on $\mathcal{I}_i$ where as $\phi_i$ is a probability mass function on $\mathcal{A}_i$.  After every time slot $T = 1,2,\dots, $ user $i$ updates its belief $\phi^i$ based on the feedback received from its receiver using Laplace estimator 
\begin{equation}
\phi^i(I_i) = \frac{T_i(I_i) + d}{T + \vert \mathcal{I}_i \vert d}, \text{ for } I_i \in \mathcal{I}_i, \label{lap_est}
\end{equation}
where $T_i(I_i)$ is the number of time instances that the interference level $I_i$ occurred up to time $T$, $\vert \mathcal{I}_i \vert$ is the cardinality of $\mathcal{I}_i$, and $d$ is any positive integer.  The Laplace estimator (\ref{lap_est}) uses Bayesian update \cite{lap_estim} and guarantees absolute continuity condition that is necessary for convergence \cite{b_learn}.  Thus, as each user $i$ plays its best response with respect to $\phi^i$, the strategies converge to an $\epsilon$-NE.\par
We will use this algorithm on some examples in the next section.
\section{Numerical Examples}\label{ne}
In this section we compare the sum rate achieved at a Nash equilibrium and a Pareto optimal point obtained by the algorithms provided above.  In all our numerical computations we choose $\delta = 0.1, \text{ and } \alpha = 0.25$ in computations of Pareto points.  In all the examples considered below, we have chosen $\tau = 0.1$ with the step size in the steepest descent method $\gamma_t = 0.5 \text{ for } t=1$ and updated after $10$ iterations as $\gamma_{t+10} = \frac{\gamma_{t}}{1+\gamma_{t}}$.  We choose a 3-user interference channel for Examples 1 and 2 below.\par
For Example 1, we take $\mathcal{H}_d = \{0.3, 1\}$ and $\mathcal{H}_c = \{0.2, 0.1\}$.  We assume that all elements of $\mathcal{H}_d, \mathcal{H}_c$ occur with equal probability, i.e., with probability 0.5.  Now, the $\tilde{H}$ matrix is not positive definite.  Thus, the algorithm in \cite{NCC15} may not converge to a NE for $\mathcal{G}_A$.  Algorithm \ref{s_min} converges to a NE not only for $\mathcal{G}_A$ but also for $\mathcal{G}_I$ and $\mathcal{G}_D$.\par
\begin{figure}
  \centering
  \includegraphics[height=5.0cm,width=10cm]{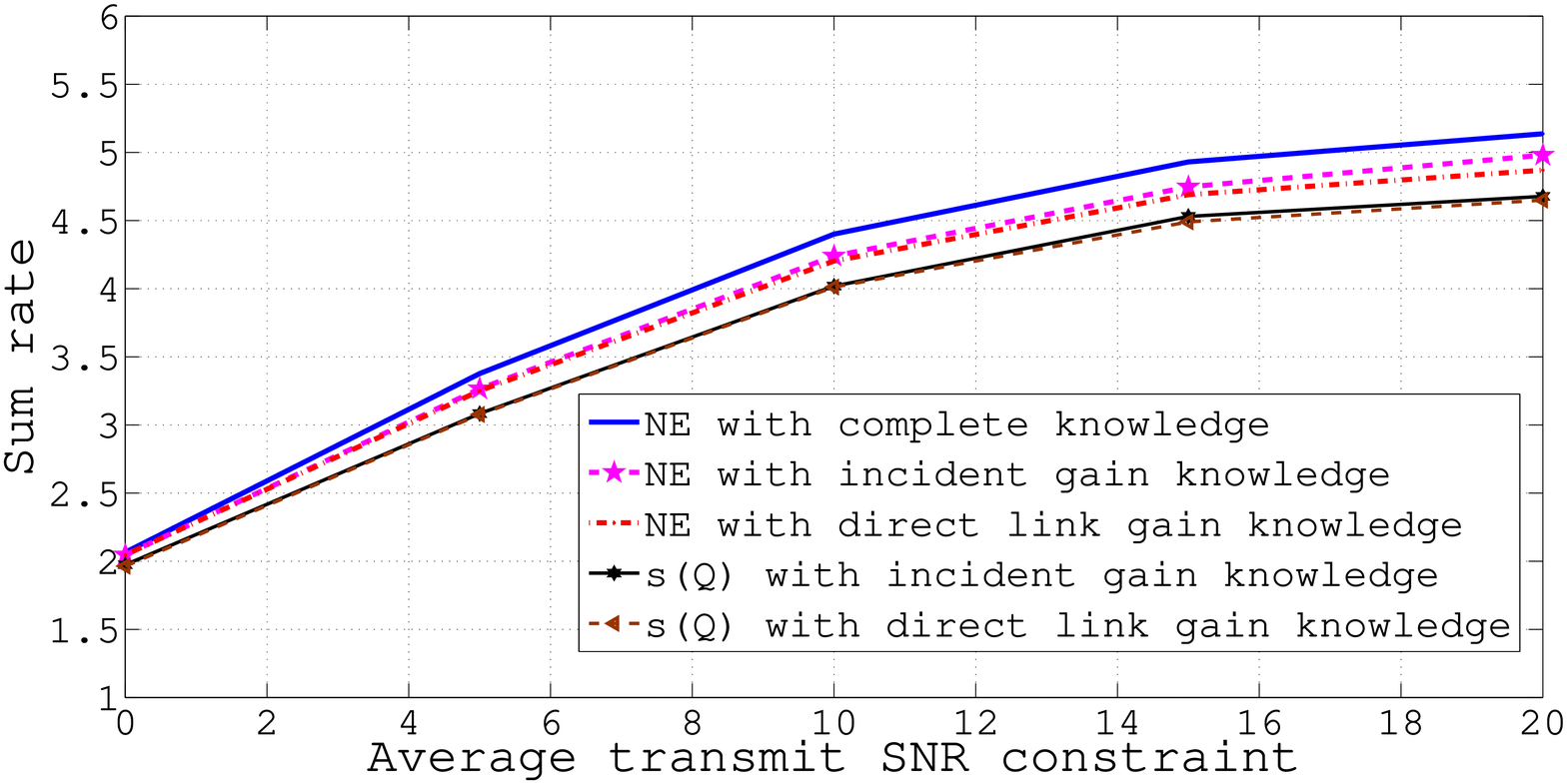}  
  \caption{Sum rate comparison at Nash equilibrium points for Example 1.}
  \label{4_plot}      
\end{figure}

We compare the sum rates for the NE under different assumptions in Figure \ref{4_plot}.  We have also computed ${\bf Q} = {\bf P}^{\dagger}$ that maximizes the corresponding lower bounds (\ref{in_lb}), evaluated the sum rate $s({\bf Q})$ and compared to the sum rate at a NE.  The sum rates at Nash equilibria for $\mathcal{G}_I$ and $\mathcal{G}_D$ are close.  This is because the values of the cross link channel gains are close and hence knowing the cross link channel gains has less impact.\par
\begin{figure}
  \centering
  \includegraphics[height=5.0cm,width=10cm]{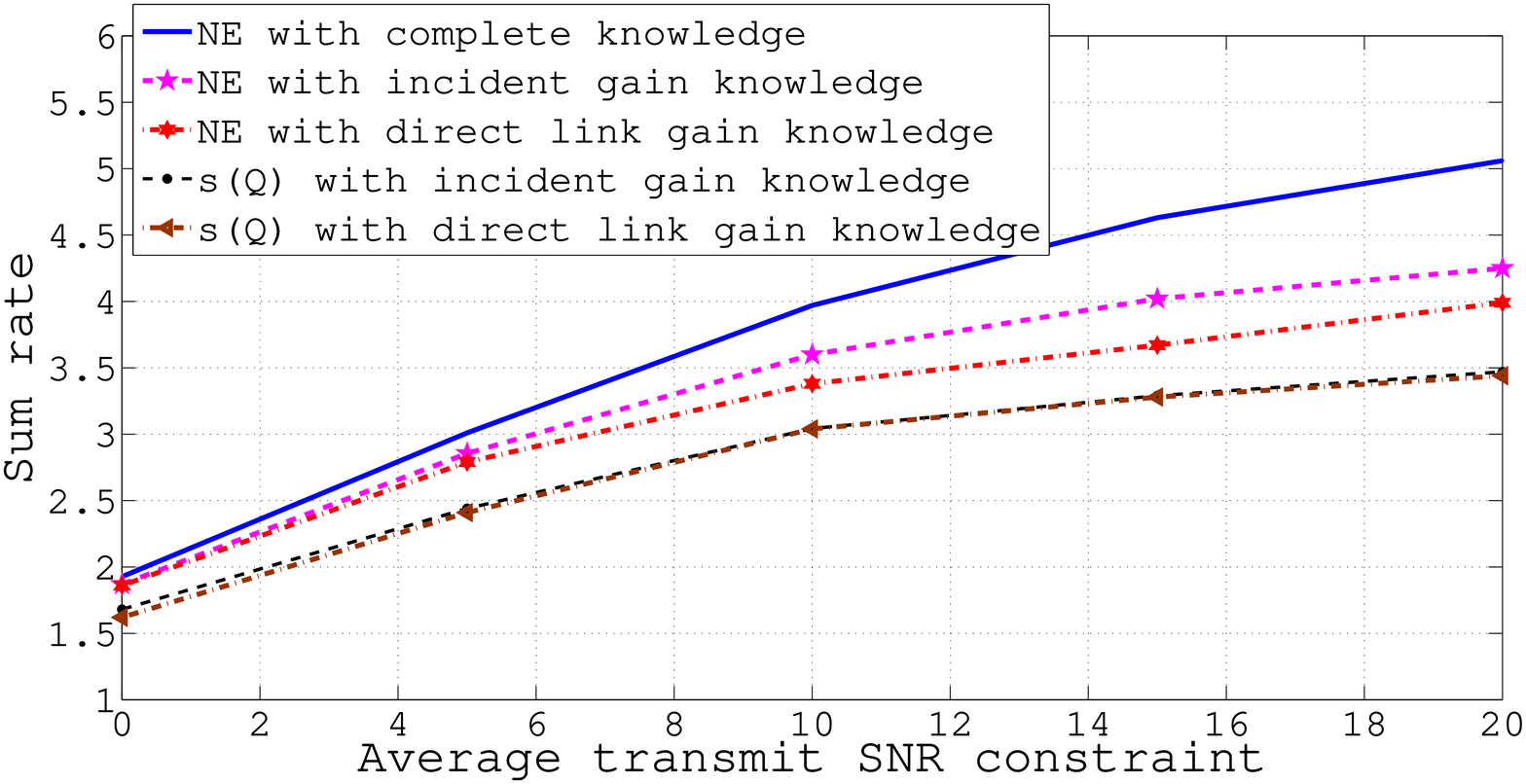}  
  \caption{Sum rate comparison at Nash equilibrium points for Example 2.}  
  \label{4_plot2}  
\end{figure}
In Example 2, we take $\mathcal{H}_d = \{0.3, 1\}$ and $\mathcal{H}_c = \{0.1, 0.5\}$.  We assume that all elements of $\mathcal{H}_d$, and $\mathcal{H}_c$ occur with equal probability.  We compare the sum rates for the NE obtained by Algorithm \ref{s_min} in Figure \ref{4_plot2}.  Now we see significant differences in the sum rates.  For this example, we compare the rates of each user at a Pareto point and Nash bargaining for games $\mathcal{G}_A$, $\mathcal{G}_I$, and $\mathcal{G}_D$ in Tables \ref{tab:g_a}, \ref{tab:g_i}, and \ref{tab:g_d} respectively.  From these tables we can observe that the Pareto optimal points yield better sum rate than at the NE.  It can also be seen that the Nash bargaining solutions provide more fairness than the Pareto points.\par
\begin{figure}
  \centering
  \includegraphics[height=5.0cm,width=10cm]{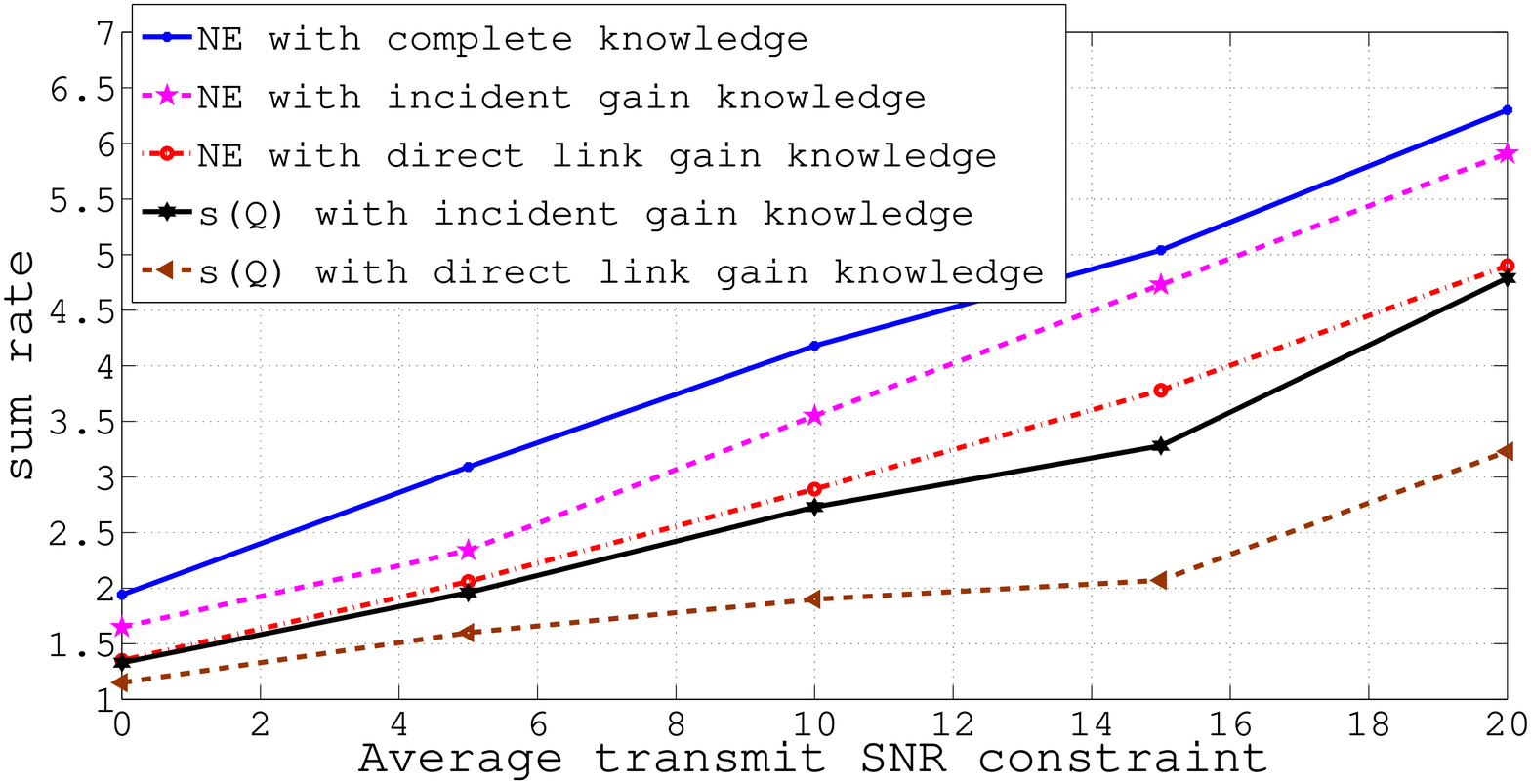}  
  \caption{Sum rate comparison at Nash equilibrium points for Example 3.}
  \label{plot3}      
\end{figure}

\begin{table}[h]
  \centering
  {\footnotesize
  \begin{tabular}{|c|c|c|}
    \hline
    SNR(dB) & Rates at Pareto point & Rates at Nash bargaining\\
    \hline
    & & \\
    0 & (0.83, 0.63, 1.02) & (0.79, 0.78, 0.81)\\ 
    & & \\    
    5 & (1.18, 1.22, 1.15) & (1.16, 1.17, 1.14)\\
    & & \\    
    10 & (1.62, 1.42, 1.82) & (1.57, 1.57, 1.57)\\
    & &\\    
    15 & (2.11, 1.90, 2.31) & (2.07, 2.05, 2.09)\\
    & &\\   
    20 & (2.54, 2.54, 2.73) & (2.45, 2.49, 2.46)\\    
    & &\\
    \hline
  \end{tabular}
  }
  \vspace{0.3cm}
  \caption[Fairness at Nash bargaining solution for $\mathcal{G}_A$]{Fairness of rates at a Pareto point and Nash bargaining for $\mathcal{G}_A$ in Example 3.}
  \label{tab:g_a}
\end{table}
\begin{table}[h]
  \centering  
  {\footnotesize
  \begin{tabular}{|c|c|c|}
    \hline 
    SNR(dB) & Rates at Pareto point & Rates at Nash bargaining\\
    \hline 
    & & \\
    0 & (0.74, 0.57, 0.95) & (0.71, 0.70, 0.72)\\ 
    & & \\    
    5 & (1.07, 1.09, 1.05) & (1.03, 1.03, 1.05)\\
    & & \\    
    10 & (1.47, 1.17, 1.68) & (1.43, 1.42, 1.43)\\
    & &\\    
    15 & (1.97, 1.67, 1.97) & (1.93, 1.95, 1.96)\\
    & &\\   
    20 & (2.38, 2.28, 2.17) & (2.32, 2.32, 2.33)\\    
    & &\\
    \hline
  \end{tabular}
  }
  \vspace{0.3cm}
  \caption[Fairness at Nash bargaining solution for $\mathcal{G}_I$]{Fairness of rates at a Pareto point and Nash bargaining for $\mathcal{G}_I$ in Example 2.}
  \label{tab:g_i}
\end{table}
\begin{table}[h]
  \centering  
  {\footnotesize
  \begin{tabular}{|c|c|c|}
    \hline 
    SNR(dB) & Rates at Pareto point & Rates at Nash bargaining\\
    \hline 
    & & \\
    0 & (0.67, 0.49, 0.82) & (0.65, 0.66, 0.64)\\ 
    & & \\    
    5 & (0.99, 0.89, 0.97) & (0.95, 0.94, 0.94)\\
    & & \\    
    10 & (1.34, 0.94, 1.54) & (1.43, 1.42, 1.43)\\
    & &\\    
    15 & (1.62, 1.37, 1.84) & (1.68, 1.72, 1.70)\\
    & &\\   
    20 & (2.24, 2.09, 2.02) & (2.20, 2.18, 2.17)\\    
    & &\\
    \hline
  \end{tabular}
  }
  \vspace{0.3cm}
  \caption[Fairness at Nash bargaining solution for $\mathcal{G}_D$]{Fairness of rates at a Pareto point and Nash bargaining for $\mathcal{G}_D$ in Example 2.}
  \label{tab:g_d}
\end{table}
We consider a 2-user interference channel in Example 3. We take $\mathcal{H}_d = \{0.1, 0.5, 1\}$ and $\mathcal{H}_c = \{0.25, 0.5, 0.75\}$.  We assume that all elements of $\mathcal{H}_d, \mathcal{H}_c$ occur with equal probability for user 1, and that the distributions of direct and cross link channel gains are identical for user 2 and are given by $\{0.1, 0.4, 0.5\}$.  In this example also, we use Algorithm \ref{s_min} to find NE for the different cases, and also obtain the lower bound for the partial information cases.  We compare the sum rates for the NE in Figure \ref{plot3}.\par
We further elaborate on the usefulness of Phase 1 in Algorithm \ref{s_min}.  We quantify the closeness of ${\bf P}$ to a NE by $g({\bf P}) = \Vert {\bf P} - T({\bf P}) \Vert$. If ${\bf P}$ is a NE then $g({\bf P}) = 0$, and for two different power allocations ${\bf P}$ and ${\bf Q}$ we say that ${\bf P}$ is closer to a NE than ${\bf Q}$ if $g({\bf P}) < g({\bf Q})$.  We now verify that the fixed point iterations in Phase 1 of Algorithm \ref{s_min} take us closer to a NE starting from any randomly chosen feasible power allocation.  For this, we have randomly generated $100$ feasible initial power allocations and run Phase 1 for $MAX = 100$ iterations for each randomly chosen initial power allocation, and compared the values of $g({\bf P})$.  In the following, we compare the mean, over the 100 initial points chosen, of the values of $g({\bf P})$ immediately after random generation of feasible power allocations, to those after running Phase 1.\par
We summarize the comparison of mean value of $g({\bf P})$ before and after Phase 1 of Algorithm \ref{s_min}, in Tables \ref{4_table1}, \ref{4_table2} and \ref{4_table3} for Examples 1, 2 and 3 respectively.  The first column of the table indicates the constrained average transmit SNR in dB.  The second and the third columns correspond to the power allocation game with complete channel knowledge, $\mathcal{G}_A$.  The fourth and the fifth columns correspond to the power allocation game with knowledge of the incident channel gains, $\mathcal{G}_I$.  The sixth and the seventh columns correspond to the power allocation game with direct link channel knowledge, $\mathcal{G}_D$.  The second, fourth and sixth columns indicate the mean of $g({\bf P})$ before running Phase 1, where ${\bf P}$ is a randomly generated feasible power allocation.  The mean value is evaluated over $100$ samples of different random feasible power allocations.  The third, fifth and seventh columns indicate the mean value of $g({\bf P})$ after running Phase 1 in Algorithm \ref{s_min} for the same random feasible power allocations.\par
It can be seen from the tables that running Phase 1 prior to Phase 2 reduces the value of $g({\bf P})$ when compared with a randomly generated feasible power allocation.  Thus, the power allocation after running Phase 1 will be a good choice of power allocation to start the steepest descent in Phase 2.  It can also be seen that for all the three examples, for $\mathcal{G}_I$ and $\mathcal{G}_D$, Phase 1 itself converges to the NE, whereas for $\mathcal{G}_A$ Phase 1 may not converge.\par
At SNR of 20dB, for $\mathcal{G}_A$, Algorithm \ref{s_min} converged in one iteration of Phase 1 and Phase 2 for Examples 1 and 3.  For Example 2, Algorithm \ref{s_min} converged after Phase 1 in the second iteration of Phase 1 and Phase 2.  Phase 2 converged to a local optimum in about 200 iterations in Example 1, about 400 iterations for Example 3 and about 250 iterations in Example 2.\par
\begin{table}[h]
  \centering
  {\footnotesize
  \begin{tabular}{|c|c|c|c|c|c|c|}
    \hline 
      & \multicolumn{2}{|c|}{$g({\bf P})$ for $\mathcal{G}_A$} & \multicolumn{2}{|c|}{$g({\bf P})$ for $\mathcal{G}_I$} & \multicolumn{2}{|c|}{$g({\bf P})$ for $\mathcal{G}_D$} \\
    \hline 
    SNR(dB) & Before Ph 1 & After Ph 1 & Before Ph 1 & After Ph 1 & Before Ph 1 & After Ph 1 \\
    \hline 
    0 & 40.82 & 8.00 $\times 10^{-4}$ & 5.01 & 0.17 $\times 10^{-4}$ & 2.48 & 0.59 $\times 10^{-14}$ \\ 
    1 & 51.39 & 0.027 & 6.42 & 0.0005 & 3.12 & 0.13 $\times 10^{-13}$\\
    5 & 96.5 & 0.15 & 11.73 & 0.0014 & 5.71 & 0.54 $\times 10^{-3}$ \\
    10 & 229.9 & 0.62 & 25.45 & 0.005 & 12.95 & 0.0023 \\
    15 & 657.3 & 2.02 & 60.6 & 0.0026 & 21.69 & 0.0027 \\
    20 & 2010.7 & 6.51 & 80.0 & 0.0029 & 31.8 & 0.0028 \\
    \hline
  \end{tabular}
  }  
  \vspace{0.3cm}
  \caption[Illustration of importance of phase 1 for Example 1]{Comparison of $g({\bf P})$ in games $\mathcal{G}_A,\mathcal{G}_I$ and $\mathcal{G}_D$ before phase 1 and after phase 1 for Example 1.}  
  \label{4_table1}
\end{table}

\begin{table}[h]
  \centering  
      {\footnotesize
  \begin{tabular}{|c|c|c|c|c|c|c|}
    \hline 
      & \multicolumn{2}{|c|}{$g({\bf P})$ for $\mathcal{G}_A$} & \multicolumn{2}{|c|}{$g({\bf P})$ for $\mathcal{G}_I$} & \multicolumn{2}{|c|}{$g({\bf P})$ for $\mathcal{G}_D$} \\
    \hline
    SNR(dB) & Before Ph 1 & After Ph 1 & Before Ph 1 & After Ph 1 & Before Ph 1 & After Ph 1 \\
    \hline
    0 & 41.68 & 0.12 & 5.14 & 0.35 $\times 10^{-4}$ & 2.47 & 0.4 $\times 10^{-15}$ \\ 
    1 & 51.43 & 0.48 & 6.40 & 0.13 $\times 10^{-3}$ & 3.17 & 0.18 $\times 10^{-14}$\\
    5 & 107.9 & 2.52 & 13.4 & 0.068 $\times 10^{-3}$ & 7.1 & 0.28 $\times 10^{-3}$ \\
    10 & 309.65 & 9.76 & 37.62 & 0.89 $\times 10^{-3}$ & 20.76 & 0.0016 \\
    15 & 948.37 & 31.68 & 98.44 & 0.0015 & 29.22 & 0.0018 \\
    20 & 2974.4 & 98.85 & 174.57 & 0.0027 & 65.15 & 0.0033 \\
    \hline
  \end{tabular}  
  }
  \vspace{0.3cm}
  \caption[Illustration of importance of phase 1 for Example 2]{Comparison of $g({\bf P})$ in games $\mathcal{G}_A,\mathcal{G}_I$ and $\mathcal{G}_D$ before phase 1 and after phase 1 for Example 2.}
  \label{4_table2}
\end{table}

\begin{table}[h]
  \centering  
  {\footnotesize
  \begin{tabular}{|c|c|c|c|c|c|c|}
    \hline 
      & \multicolumn{2}{|c|}{$g({\bf P})$ for $\mathcal{G}_A$} & \multicolumn{2}{|c|}{$g({\bf P})$ for $\mathcal{G}_I$} & \multicolumn{2}{|c|}{$g({\bf P})$ for $\mathcal{G}_D$} \\
    \hline
    SNR(dB) & Before Ph 1 & After Ph 1 & Before Ph 1 & After Ph 1 & Before Ph 1 & After Ph 1 \\
    \hline
    0 & 12.30 & 0.04 & 4.07 & 0.95 $\times 10^{-5}$ & 2.30 & 0.42 $\times 10^{-4}$ \\ 
    1 & 14.82 & 0.05 & 4.81 & 0.22 $\times 10^{-4}$ & 2.80 & 0.93 $\times 10^{-4}$\\
    5 & 34.21 & 0.28 & 10.90 & 0.47 $\times 10^{-3}$ & 5.71 & 0.89 $\times 10^{-3}$\\
    10 & 104.74 & 0.89 & 32.34 & 0.0014 & 16.82 & 0.0007 \\
    15 & 325.75 & 2.43 & 103.72 & 0.0016 & 44.72 & 0.001 \\
    20 & 1010.10 & 9.27 & 271.46 & 0.0017 & 107.96 & 0.002 \\
    \hline
  \end{tabular} 
  }
  \vspace{0.3cm}
  \caption[Illustration of importance of phase 1 for Example 3]{Comparison of $g({\bf P})$ in games $\mathcal{G}_A,\mathcal{G}_I$ and $\mathcal{G}_D$ before phase 1 and after phase 1 for Example 3.}
  \label{4_table3}
\end{table}
We have run Algorithm \ref{s_min} on many more examples and found that for $\mathcal{G}_I$ and $\mathcal{G}_D$, Phase 1 itself converged to the NE.\par
\begin{figure}
  \centering
  \includegraphics[height=5.0cm,width=10cm]{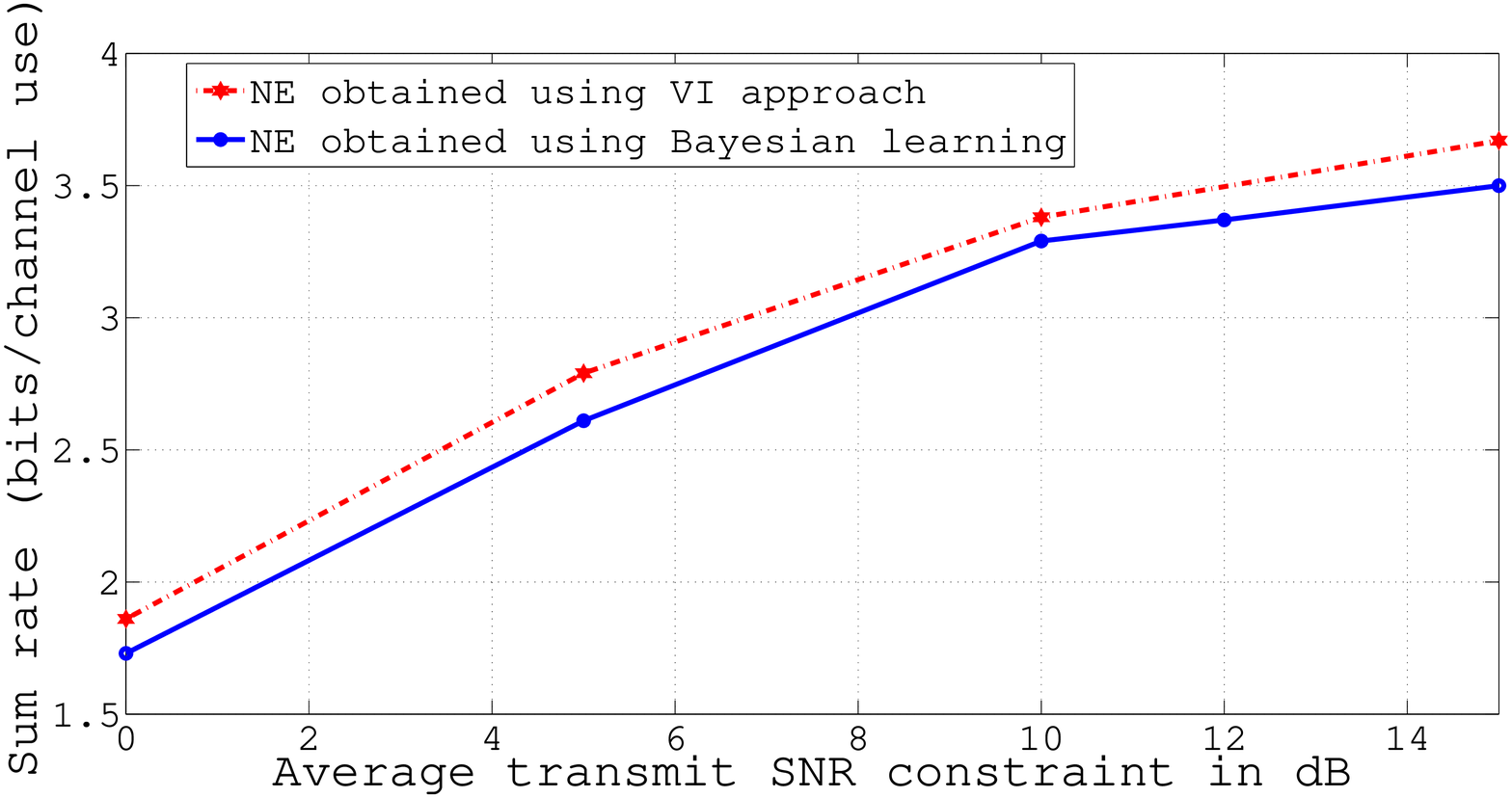}
  \caption[Comparison of heuristic and Bayesian learning]{Sum rate comparison at Nash equilibrium points of $\mathcal{G}_D$ using VI approach and Bayesian learning for Example 2.}
  \label{plot_baye}      
\end{figure}

We illustrate Bayesian learning for Example 2 with $\mathcal{H}_d^{(i)} = \{0.3, 1\}$ and $\mathcal{H}_c^{(i)} = \{0.5, 0.1\}$ for each $i = 1,2,3$.  We assume that all elements of $\mathcal{H}_d$, and $\mathcal{H}_c$ occur with equal probability.  Each user transmits data at rate $r_i^{(D)}$ given by (\ref{rate_direct}) with a power level between $0$ and $50$ which is a multiple of $5$.  Each user has a belief on the distribution of interference experienced by its receiver and uses Bayesian learning to find a NE of $\mathcal{G}_D$.  We tabulate these rates in Table \ref{table4} for all the three users at a $\epsilon$-NE computed via the Bayesian learning algorithm, and we also compare the sum rates at NE obtained using variational inequality approach and Bayesian learning in Figure \ref{plot_baye}.  It can be seen from Figure \ref{plot_baye} that the sum rates achieved via the VI heuristic and using Bayesian learning are very close.\par
In our simulations for Example 2, At transmit SNR of 15dB, to find a NE of $\mathcal{G}_I,\text{ }\mathcal{G}_D$ using variational inequality approach requires about $150$ iterations in phase $1$.  The overall run time of the VI based heuristic algorithm is about $71.5761$ seconds for $\mathcal{G}_I$ and $65.9362$ seconds for $\mathcal{G}_D$ on an \emph{i5-2400} processor with clock speed $3.10$GHz.  Bayesian learning converges to a NE in about $10,000$ iterations and the run time for the program on the same processor is about $82.7934$ seconds.  Even though Bayesian learning requires a larger number of iterations to converge, its per iteration complexity is less which reduces run time.  But this run time increases significantly if we increase the number of feasible power levels.

\begin{table}[h]
  \centering  
  {\footnotesize
  \begin{tabular}{|c|c|}
    \hline 
    SNR(dB) & Rates of users at NE \\
    \hline
    & \\
    0 & (0.59, 0.56, 0.58) \\
    & \\    
    5 & (0.87, 0.86, 0.88)\\
    & \\    
    10 & (1.09, 1.10, 1.10)\\
    & \\    
    12 & (1.13, 1.12, 1.12)\\
    & \\   
    15 & (1.16, 1.17, 1.17) \\    
    & \\
    \hline
  \end{tabular}
  }
  \vspace{0.3cm}
  \caption{Rates of users at $\epsilon$-NE from Bayesian learning.}
  \label{table4}
\end{table}
\section{Conclusions}\label{concl}
We have considered a channel shared by multiple transmitter-receiver pairs causing interference to one another.  We formulated stochastic games for this system in which transmitter-receiver pairs may or may not have information about other pairs' channel gains.  Exploiting variational inequalities, we presented a heuristic algorithm that obtains a NE in the various examples studied, quite efficiently.\par
In the games with partial information, we presented a lower bound on the utility of each user at any NE.  A utility of at least this lower bound can be attained by a user using a water-filling like power allocation, that can be computed with the knowledge of the distribution of its own channel gains and of the average power constraints of all the users.  This power allocation is especially useful when any transmitter fails to receive the power variables from the other transmitters that are required for it to compute its NE power allocation.\par
In all the games, i.e., $\mathcal{G}_A$, $\mathcal{G}_I$ and $\mathcal{G}_D$, we also provide algorithms to compute the Pareto points and Nash Bargaining solutions which yield better sum rate than the NE.  The Nash Bargaining solutions are fairer to users than the Pareto points.  Bayesian learning has been used to compute NE for general channel conditions.  It is observed that, even though Bayesian learning takes more iterations to compute NE than the heuristic, it requires less information about the other users and their strategies.  But to use Bayesian learning, we quantize the power levels and it is the price we pay for not having more information.
\section*{Acknowledgement}
This work is partially supported by funding from ANRC.

\end{document}